\documentclass[preprint,prd,nofootinbib,tightenlines,amsmath]{revtex4}
\usepackage{bbm}
\usepackage{amsfonts}
\usepackage{mathrsfs}
\usepackage{graphicx}
\usepackage{bm}

\begin{document}
\baselineskip=15pt \parskip=5pt

\vspace*{3em}

\title{Searching for New Physics in $D^0\rightarrow \mu^+\mu^-,\; e^+e^-, \;\mu^{\pm}e^{\mp}$ at BES and/or Super Charm-Tau Factory}

\author{Lian-Bao Jia$^1$}
\email{jialb@mail.nankai.edu.cn}
\author{Ming-Gang Zhao$^1$}
\author{Hong-Wei Ke$^2$}
\author{Xue-Qian Li$^1$}

\affiliation{1. School of Physics, Nankai University, Tianjin
300071, China \\
2. School of Physics, Tianjin University, Tianjin, 300072, China\\}

\begin{abstract}

In contrast with $B^0-\bar B^0$, $B_s-\bar B_s$ mixing where the
standard model (SM) contributions overwhelm that of new physics
beyond standard model (BSM), a measured relatively large $D^0-\bar D^0$ mixing
where the SM contribution is negligible, definitely implies the
existence of new physics BSM. It is natural to consider that the
rare decays of D meson might be more sensitive to new physics, and
the rare decay $D^0\to \mu^+\mu^-$ could be an ideal area to search
for new physics because it is a flavor changing process. In this
work we look for a trace of new physics BSM in the leptonic decays of $D^0$, concretely we discuss the contributions of unparticle or an extra
gauge boson $Z'$ while imposing the constraints set by fitting the $D^0-\bar D^0$ mixing data. We find that the long-distance SM effects for $D^0\to l\bar l$ still exceed those
contributions of the BSM under consideration, but for a double-flavor changing
process such as $D^0\to \mu^{\pm}e^{\mp}$, the new physics
contribution would be significant.

\end{abstract}

\maketitle

\section{Introduction}

One of the main goals of the study on lower energy processes is to
look for traces of new physics BSM and it is
mutually complementary  with the very high energy processes at
LHC. It is believed that the SM is very successful that its
predictions are well consistent with all the present experimental
data. But the SM is still an effective theory. The consistency is
because at lower energy scales the contributions from new physics
BSM are much smaller than that of SM which dominates all the
processes. Even though the effects of new physics are small, they
may manifest in some precise measurements and leave traces.
Generally, BSM effects may show up at rare processes where the SM
contributions are forbidden  or strongly suppressed. Therefore more
theorists and experimentalists have growing interests in the rare
decays of heavy flavor mesons and baryons. Such studies may find
traces of BSM and provide valuable information to LHC for designing
new experiments.

As is well understood, the SM dominates the $B^0-\bar B^0$ and
$B_s-\bar B_s$ mixing due to an enhancement factor $m_t^2/M_W^2$ in
the box-diagrams, thus contributions of new physics BSM are much
smaller than that of SM. By contrary, the SM contributions to
$D^0-\bar D^0$ mixing are negligible because the intermediate quarks
in the box are b and s which are much lighter than $M_W$. The first
evidence of $D^0-\bar{D}^0$ oscillation is presented by the BaBar~\cite{Aubert:2007wf} and Belle \cite{Staric:2007dt} Collaborations
and later further confirmed by the CDF
Collaboration~\cite{Aaltonen:2007ac} in 2007. The relatively large
mixing implies the existence of new physics BSM. There have been
many models which offer a flavor-changing-neutral current (FCNC)
and enhance the mixing to the observational level. For example, the
Littlest Higgs Model~\cite{Chen:2007yn}, the fourth
generation~\cite{Hou:2006mx}, non-universal $Z'$~\cite{He:2007iu}
and unparticle~\cite{Li:2007by} etc., can result in a larger
$D^0-\bar{D}^0$ mixing. Thus it motivates people to look for rare
decay processes where the SM contributions are suppressed, so that
the new physics effects would not be buried in the SM background.
Taking into account the constraints set by $D^0-\bar{D}^0$ mixing,
we turn to investigate new physics contributions to the rare decays
$D^0\rightarrow l'^+ l^-$.

Recently, an intensive study on the leptonic decays of $B^0(\bar
B^0)$ and $B_s(\bar B_s)$ is carried out. It seems that no evidence
of new physics BSM is needed to explain the present data obtained by
LHCb~\cite{Aaij:2012nna,Aaij:2013aka} and
CMS~\cite{Chatrchyan:2013bka}. One may wonder if $D$ is more
sensitive to new physics as it happens to the $D^0-\bar D^0$ mixing. As
the existence of a flavor changing neutral current can explain the
$D^0-\bar D^0$ mixing, the same mechanism should apply to the
leptonic decays $D^0\to \mu^+\mu^-,\; e^+e^-$, and it might cause
sizable effects to enhance the rates of the leptonic decays.
Definitely, such mechanism would also apply to leptonic decays of
$B^0$ and $B_s$ even though they do not manifest for the $B-\bar B$
mixing.

In this work we calculate the decay rates of $D^0\to \mu^+\mu^-,\;
e^+e^-$ in terms of both the unparticle model and an extra gauge
boson $Z'$. Our numerical results indicate that the contributions of
the new physics of concern to the decay rates do not exceed that
coming from the long-distance SM effects. But it is not the end of the
story, as we proceed to study the lepton-flavor changing decay
$D^0\to \mu^+e^-(\mu^-e^+)$ which is double-flavor changing process,
the new physics may be significant. Moreover, when we consider the
possible CP violation, the role of new physics might also be
important.

Our strategy is that we employ the model parameters obtained by
fitting the data of $D^0-\bar D^0$ mixing for both unparticle and
extra gauge boson $Z'$ scenarios, then apply them to estimate the
decay rates under consideration.

This work is organized as follows: after this short introduction, we
formulate the decay rates of $D^0\to \mu^+\mu^-,\; e^+e^-$  and
$\mu^{\pm}e^{\mp}$ in section II. In section III, we present our
numerical results along with all the input model parameters. In
section IV, we discuss possible measurement schemes on the leptonic
decays and the lepton-flavor-changing decay, and the last section is
devoted to our conclusion and a brief discussion.

\section{Contributions of new physics BSM to $D^0\rightarrow l'^+ l^-$}

The SM contribution to the decay of $D^0\rightarrow l'^+ l^-$ has been
estimated as the short distance contribution to $\mathcal
{B}_{D^0\rightarrow \mu^+ \mu^-}$ is of order
$10^{-19}\sim10^{-18}$~\cite{Gorn:1978sb,Pakvasa:1994ni,Burdman:2001tf},
while taking into the long distance contributions, the branching
ratio can reach a level of
$10^{-13}$~\cite{Pakvasa:1994ni,Burdman:2001tf}. The branching ratio
of ${D^0\rightarrow e^+ e^-}$ is of order
$10^{-23}$~\cite{Burdman:2001tf}, and the decay mode $D^0\rightarrow
\mu^{\pm} e^{\mp}$ is deeply suppressed in SM. Obviously, these
rates are too small to be detected by the present facilities. Our
goal of this work is to investigate if the new physics BSM would
result in larger rates for those decays. In this work we only let
ourselves concentrate on two possible models: unparticle
\cite{Georgi:2007ek} and non-universal boson
$Z'$~\cite{Barger:2003hg,He:2004it,Cheung:2006tm,He:2006bk,Chiang:2006we,Baek:2006bv}.
These models have been thoroughly discussed in literature, so that
first we briefly show how to extract model parameters from
$D^0-\bar{D}^0$ mixing data, then we formulate the new physics
contributions to the rare decays $D^0\rightarrow l'^+ l^-$.

\subsection{Determination of the new physics parameters by fitting $D^0-\bar{D}^0$ mixing}

A detectable $D^0-\bar{D}^0$ mixing has been measured, but as
indicated the SM contribution cannot induce a detectable mixing. There are
two crucial parameters for the $D^0-\bar{D}^0$ mixing which are
experimentally measured via the $D^0-\bar{D}^0$ oscillation. The
physical eigen-states are
\begin{eqnarray}
\mid D_{1,2}\rangle =p\mid D^0 \rangle~\pm~q \mid \bar{D}^0 \rangle,
\end{eqnarray}
and the measurable parameters  $x, y$ are defined
as $
x\equiv \frac{m_{1}-m_{2}}{\Gamma} =\frac{\Delta m_D}{\Gamma}$
and
$ y\equiv \frac{\Gamma_{1}-\Gamma_{2}}{2 \Gamma}
=\frac{\Delta\,\Gamma_D}{2 \Gamma}$,
where $\Gamma=(\Gamma_{1}+\Gamma_{2})/2$. Experimentally, the
``rotated" parameters $x'$, $y'$ are also used (for more details,
see, e.g.,~\cite{Li:2010af}). The updated Belle
results~\cite{Peng:2013wsa&Li:2013mnl} are $x=(0.56 \pm
0.19^{+0.03+0.06}_{-0.09-0.09})\%$, $y=(0.30 \pm
0.15^{+0.04+0.03}_{-0.05-0.06})\%$. No evidence of CP violation was
observed at Belle so far and it is consistent with the observed
results at LHCb~\cite{Aaij:2013wda,Aaij:2013ria}.

\subsubsection{Constraints on the parameters of unparticle scenario}

The scale invariant unparticle scenario was proposed by
Georgi~\cite{Georgi:2007ek}, which has a non-integral scale
dimension $d_{\mathcal {U}}$ below a typical energy scale
$\Lambda_{\mathcal {U}}$. In the scenario of unparticle, different
flavors can be coupled to unparticle, so that FCNC can be induced at tree
level. The scalar, vector unparticle fields are denoted as
$O_{\mathcal {U}}$, $O_{\mathcal {U}}^{\mu}$. The propagator of
scalar unparticle is~\cite{Georgi:2007si,Cheung:2007zza,Luo:2007bq}
\begin{eqnarray}
\int\,d^{\,4} x e^{iP.x} \langle0\mid T O_{\mathcal {U}} (x)
O_{\mathcal {U}} (0) \mid 0
\rangle\nonumber\\
=i\frac{A_{d_{\mathcal {U}}}}{2 \sin (d_{\mathcal {U}}\pi)
}\frac{1}{(P^2+i\epsilon)^{2-d_{\mathcal {U}}}}e^{-i(d_{\mathcal
{U}}-2)\pi},
\end{eqnarray}
where $A_{d_{\mathcal {U}}}$ is
\begin{eqnarray}
A_{d_{\mathcal {U}}}=\frac{16\pi^{5/2}}{(2\pi)^{2d_{\mathcal
{U}}}}\frac{\Gamma(d_{\mathcal {U}}+1/2)}{\Gamma(d_{\mathcal
{U}}-1)\Gamma(2d_{\mathcal {U}})}.
\end{eqnarray}
The vector unparticle propagator is~\cite{Grinstein:2008qk}
\begin{eqnarray}
&&\int\,d^{\,4} x e^{iP.x} \langle0\mid T O_{\mathcal {U}}^{\mu} (x)
O_{\mathcal {U}}^{\nu} (0) \mid 0
\rangle\,\,\,\nonumber\\
&&=i\frac{A_{d_{\mathcal {U}}}}{2 \sin (d_{\mathcal {U}}\pi)
}\frac{-g^{\mu\nu}{+\frac{2(d_{\mathcal {U}}-2)}{d_{\mathcal
{U}}-1}}\frac{P^{\mu}P^{\nu}}{P^2}}{(P^2+i\epsilon)^{2-d_{\mathcal
{U}}}}e^{-i(d_{\mathcal {U}}-2)\pi}.
\end{eqnarray}
The unitarity bounds on the non-integral scale dimension
$d_{\mathcal {U}}$ below the typical energy scale $\Lambda_{\mathcal
{U}}$ are that $d_{\mathcal {U}}\geq1$ for scalar unparticle and
$d_{\mathcal {U}}\geq3$ for vector
unparticle~\cite{Grinstein:2008qk}.

The mass and width differences are related to the mixing elements,
$\Delta m_D = 2 \mid M_{12} {\mid}$ and $\Delta \Gamma_D = 2 \mid
\Gamma_{12} {\mid}$. In the case of CP conservation, the scalar
unparticle's contribution to the mass difference (for more, see
e.g.~\cite{Li:2007by,Chen:2007cz}) is
\begin{eqnarray}
\Delta m^{\mathcal {U}}_{D}=\frac{5}{3}\frac{f^2_D \hat{B}_D}{m_D
}\frac{A_{d_{\mathcal {U}}}}{4}(\frac{m_D}{\Lambda_{\mathcal
{U}}})^{2 d_{\mathcal {U}}}(\frac{m_D}{m_c})^2 \mid c_S^{\,uc}
{\mid}^2 \mid \cot d_{\mathcal {U}}\pi \mid .\label{s-unp}
\end{eqnarray}
For vector unparticle, the result is
\begin{eqnarray}
\Delta m^{\mathcal {U}}_{D}=\frac{f^2_D \hat{B}_D}{m_D
}\frac{A_{d_{\mathcal {U}}}}{4}(\frac{m_D}{\Lambda_{\mathcal
{U}}})^{2 d_{\mathcal {U}}-2}\mid c_V^{\,uc}
{\mid}^2[\frac{8}{3}-\frac{2(d_{\mathcal {U}}-2)}{d_{\mathcal
{U}}-1}\frac{5}{3}(\frac{m_D}{m_c})^2]\mid\cot d_{\mathcal
{U}}\pi\mid.\label{v-unp}
\end{eqnarray}
Here the Wick contraction factors have been taken into
consideration. $f_D$ is the decay constant, $f_D \simeq$0.2 GeV, and
$\hat{B}_D$ is a factor related to non-perturbative QCD with order
of unity, $\hat{B}_D\simeq 1$ corresponding to the vacuum saturation~\cite{Burdman:2003rs}.
$m_D$ is $D^0$ meson mass, and $\Lambda_{\mathcal {U}}$ is of order
TeV. $c_S$, $c_V$ are the coupling parameters.

For the mixing induced by unparticle, the relation
$\Gamma_{12}^{\mathcal {U}}/2=M_{12}^{\mathcal {U}} \tan
(d_{\mathcal {U}}\pi)$ holds, as given in Ref.~\cite{Chen:2007cz}.
Thus $\Delta \Gamma^{\mathcal {U}}_{D}/2=\Delta m^{\mathcal
{U}}_{D}\mid \tan (d_{\mathcal {U}}\pi)\mid$. As the
contributions to the mass and width differences are totally from
unparticle, i.e. ignoring the contributions from the SM and other BSMs, $\Delta m^{\mathcal {U}}_{D}\sim\Delta m_D$ and $\Delta
\Gamma^{\mathcal {U}}_{D}\sim\Delta \Gamma_D$. The measurement
values of x, y can be used to determine
the unparticle parameters and then applied to calculate the rates of
$D^0\rightarrow l'^+ l^-$.

\subsubsection{Constraints on the parameters of the non-universal $Z'$}

Instead of unparticle scenario, let us turn to another possible BSM. In this scenario,
a tree-level FCNC is induced by the new non-universal gauge
boson $Z'$. Some phenomenological applications of the non-universal
$Z'$ have been widely
studied~\cite{Barger:2003hg,He:2004it,Cheung:2006tm,He:2006bk,Chiang:2006we,Baek:2006bv}.
It was applied to the $D^0-\bar{D}^0$ mixing by the authors of \cite{He:2007iu}.
The flavor-changing couplings of $Z'$ to quarks and
leptons are in the form
\begin{eqnarray}\label{mixing}
\mathcal {L}&=&\frac{g}{2}\tan \theta_W (\tan \theta_R+ \cot
\theta_R)(\sin \xi_Z Z_{\mu}+\cos \xi_Z Z'_{\mu})\nonumber\\
&&\times(V^{d\ast}_{R\,bi}
V^{d}_{R\,bj}\bar{d}_{R\,i}\Gamma^{\mu}{d}_{R\,j}-V^{u\ast}_{R\,ti}
V^{u}_{R\,tj}\bar{u}_{R\,i}\Gamma^{\mu}{u}_{R\,j}+\bar{\tau}_{R}\Gamma^{\mu}\tau_{R}-\bar{\nu}_{R\tau}\Gamma^{\mu}\nu_{R\tau})\,,
\end{eqnarray}
where g is the $SU(2)_L$ coupling, and $\theta_W$ is the Weinberg
angle, as in SM. $\theta_R$ is related to the right-handed
interaction strength, and $\xi_Z$ parameterizes the $Z-Z'$ mixing
angle. $V^{u,d}_{R\,ij}$ are the matrix rotating the right-handed
up(down)-type quarks from their weak eigen-states to their mass
eigen-states.

The bound set by the LEP-II measurements can be approximated in a relation
form~\cite{He:2002ha,He:2003qv}, $\tan \theta_W \cot \theta_R
\frac{M_W}{M_{Z'}}\sim1$. Supposing that the measured $x$ is fully
determined by the contribution of $Z'$, the $D^0-\bar{D}^0$ mixing
constrains the matrix element $\mid V^{u\ast}_{R\,tu}
V^{u}_{R\,tc} \mid$ is~\cite{He:2007iu}
\begin{eqnarray}
\mid V^{u\ast}_{R\,tu} V^{u}_{R\,tc} \mid \lesssim 2.0 \times
10^{-4} \,.
\end{eqnarray}
This bound will used for evaluating the rates of $D^0\rightarrow
l'^+ l^-$ decays.

\subsection{Unparticle contribution to $D^0\rightarrow l'^+ l^-$}

\begin{figure}[t]
\includegraphics[width=2.4in]{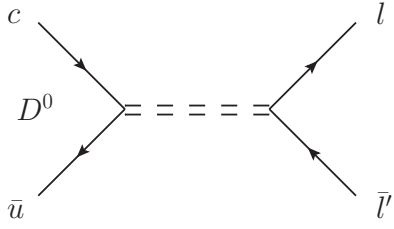} \vspace*{-1ex}
\caption{Unparticle induced $D^0\rightarrow l'^+ l^-$
decays.}\label{dL}
\end{figure}

For the mixing, as shown in Eqs.(\ref{s-unp}),(\ref{v-unp}), the
vector unparticle's contribution is more suppressed by a factor
$(\frac{m_D}{\Lambda_{\mathcal {U}}})^{2d_{\mathcal {U}}}$ compared
with the scalar unparticle.  The unparticle effect
on $B_s\rightarrow \mu^+ \mu^-$ was discussed in Ref.~\cite{Lee:2013dma}. The leptonic decay $D^0\rightarrow l'^+ l^-$ is similar. Therefore,
here we just consider the scalar unparticle contribution, and the Feynman diagram
is presented in Fig.\ref{dL}.
The effective interaction of scalar unparticle with quarks and/or leptons
is
\begin{eqnarray}
\frac{c_S^{\,{q}'q}}{\Lambda_{\mathcal {U}}^{d_{\mathcal
{U}}}}\bar{q}'\gamma_{\mu}(1-\gamma_5)q\partial^{\mu}O_{\mathcal
{U}}+\frac{c_S^{\,{l}'l}}{\Lambda_{\mathcal {U}}^{d_{\mathcal
{U}}}}\bar{l}'\gamma_{\mu}(1-\gamma_5)l \partial^{\mu}O_{\mathcal
{U}}+h.c.\,,
\end{eqnarray}
where $c_S^{\,{q}'q}$, $c_S^{\,{l}'l}$ are the coupling constants for
quarks and leptons respectively.

Including contributions of SM and unparticle, the decay width of $D^0\rightarrow
l'^+ l^-$ is
\begin{eqnarray}
\Gamma_{D^0\rightarrow l'^+ l^-}=\frac{1}{16 \pi m_D}\beta_f\mid
\langle l'^+ l^-\mid \mathcal {H}_{SM} + \mathcal {H}_{\mathcal
{U}}\mid D^0\rangle\mid^2\,,
\end{eqnarray}
where
\begin{eqnarray}
\beta_f=\sqrt{1-\frac{2(m_{l'}^2+m_{l}^2)}{m_D^2}+\frac{(m_{l'}^2-m_{l}^2)^2}{m_D^4}}\,,
\end{eqnarray}
$\mathcal {H}_{SM}$, $\mathcal {H}_{\mathcal {U}}$ are SM and
unparticle Hamiltonians respectively. $\mathcal
{H}_{\mathcal {U}}$ is in the form
\begin{eqnarray}
\mathcal {H}_{\mathcal {U}}&=&-\frac{A_{d_{\mathcal {U}}}}{2 \sin
(d_{\mathcal {U}}\pi) }\frac{1}{m_D^4}e^{-i(d_{\mathcal
{U}}-2)\pi}(\frac{m_D^2}{\Lambda_{\mathcal
{U}}^2})^{d_{\mathcal {U}}}c_S^{\,uc} c_S^{\,{l}'l}[m_{l}\bar{l}(1-\gamma_5)l'-m_{l'}\bar{l}(1+\gamma_5)l']\nonumber\\
&&\times[m_u\bar{u}(1-\gamma_5)c-m_c\bar{u}(1+\gamma_5)c]\,,
\end{eqnarray}
where the relation $P^2= m_D^2$ has been used. Let us first
consider only the unparticle contribution to the decay rate. The decay
width is
\begin{eqnarray}
\Gamma_{D^0\rightarrow l'^+ l^-}^{\,\mathcal {U}}&=&\frac{1}{16 \pi
m_D}\beta_f\mid \frac{A_{d_{\mathcal {U}}}}{2 \sin (d_{\mathcal
{U}}\pi) }\frac{1}{m_D^4}e^{-i(d_{\mathcal
{U}}-2)\pi}(\frac{m_D^2}{\Lambda_{\mathcal {U}}^2})^{d_{\mathcal
{U}}}c_S^{\,uc} c_S^{\,{l}'l}\mid^2\nonumber\\
&&\times[4(m_l'^2+m_l^2)(m_D^2-m_l'^2-m_l^2)+16m_l'^2 m_l^2]f_D^2
m_D^4\,.\label{un}
\end{eqnarray}
Taking $\Delta m^{\mathcal {U}}_{D}$,$\Delta \Gamma^{\mathcal
{U}}_{D}$ into the above formula (\ref{un}), we have
\begin{eqnarray}
\Gamma_{D^0\rightarrow l'^+ l^-}^{\,\mathcal {U}}&=&\frac{1}{16 \pi
m_D}\beta_f \mid \frac{c_S^{\,{l}'l}}{c_S^{\,uc}}\mid^2
[(\frac{6}{5}\Delta m^{\mathcal {U}}_{D})^2+
(\frac{6}{5}\frac{\Delta
\Gamma_{D}^{\mathcal {U}}}{2})^2]\frac{1}{f_D^2 m_D^2}(\frac{m_c}{m_D})^4\nonumber\\
&&\times[4(m_l'^2+m_l^2)(m_D^2-m_l'^2-m_l^2)+16m_l'^2 m_l^2]\,.
\end{eqnarray}

\subsection{Non-universal $Z'$ contribution to $D^0\rightarrow l'^+ l^-$}

In the limit of the mixing angle $\xi_Z\sim0$, we only consider the
contribution of $Z'$. The decay width of $D^0\rightarrow\mu^+ \mu^-$
can be formulated as ~\cite{He:2007iu}
\begin{eqnarray}
\Gamma_{D^0\rightarrow \mu^+ \mu^-}&\approx& \frac{G_F^{\,2} m_D
m_{\mu}^2 f^2_D}{16\pi}\mid V^{u\ast}_{R\,tu} V^{u}_{R\,tc} \mid^2
(\tan \theta_W \cot \theta_R \frac{M_W}{M_{Z'}})^4 \tan^4 \theta_R
\,.
\end{eqnarray}
For the process $D^0\rightarrow e^+ e^-$, the decay width is
proportional to the lepton mass square, so it is suppressed compared to
$D^0\rightarrow \mu^+ \mu^-$.

In formula (\ref{mixing}), the rotation only applies to the quark
sector, one may naturally generalize the lagrangian to involve a
rotation at the lepton sector.   The lagrangian can be re-written as
\begin{eqnarray}
\mathcal {L}&=&\frac{g}{2} \tan \theta_W (\tan \theta_R+ \cot
\theta_R)(\sin \xi_Z Z_{\mu}+\cos \xi_Z Z'_{\mu})\nonumber\\
&&\times(V^{l^c\ast}_{R\,\tau i} V^{l^c}_{R\,\tau
j}\bar{\tau}_{R\,i}\Gamma^{\mu}{\tau}_{R\,j}-V^{\nu\ast}_{R\,\tau i}
V^{\nu}_{R\,\tau j}\bar{\nu}_{R\,i}\Gamma^{\mu}{\nu}_{R\,j})\,,
\end{eqnarray}
where $V^{l^c,\nu}_{R\,ij}$ are matrix elements rotating the lepton
weak eigen-states to the mass eigen-states, moreover, this
lagrangian allows flavor changes as $i$ is not necessary to be equal
to $j$. In this case, the lepton-flavor-changing interaction induced
by $Z'$ would occur at tree level.  The decay width of
$D^0\rightarrow \mu^+ e^-$ can be obtained,
\begin{eqnarray}
\Gamma_{D^0\rightarrow \mu^+ e^-}\approx \frac{G_F^{\,2} m_D
m_{\mu}^2 f^2_D}{32\pi}\mid V^{u\ast}_{R\,tu} V^{u}_{R\,tc} \mid^2
\mid V^{l^c\ast}_{R\,\tau e} V^{l^c}_{R\,\tau \mu} \mid^2 (\tan
\theta_W \cot \theta_R \frac{M_W}{M_{Z'}})^4 \,.
\end{eqnarray}
In the following computations, we are simply going to employ the model parameters obtained
by others and will list them in next section.

\section{Numeral analysis on $D^0\rightarrow l'^+ l^-$}

In the following, we present our numeral results of the decay
$D^0\rightarrow l'^+ l^-$ based on the new physics BSM, both
unparticle and non-universal $Z'$.

\subsection{Unparticle}

First we discuss the unparticle contribution to the decays
$D^0\rightarrow l'^+ l^-$. Relevant parameters are input as
$m_{c}=1.275\pm0.025$ GeV, $m_{D}=1.86486\pm0.00013$ GeV, and the
mean lifetime of $D^0$ meson $(410.1\pm1.5)\times 10^{-15}$
s~\cite{Beringer:1900zz}. The updated Belle
results~\cite{Peng:2013wsa&Li:2013mnl} are used to constrain the new
physics contributions, taking the central values, $x\sim0.056$,
$y\sim0.030$, and $x^2 +y^2\sim4.0\times10^{-5}$. Though with a
large uncertainty of $x^2 +y^2$, it should be taken as an upper bound
of unparticle contribution. The branching ratios $\mathcal
{B}^{\mathcal {U}}_{D^0\rightarrow l'^+ l^-}$ with the contributions
from only unparticle are
\begin{eqnarray}
\mathcal {B}^{\mathcal {U}}_{D^0\rightarrow \mu^+
\mu^-}\lesssim4.8\times10^{-19}\mid \frac{c_S^{\,\mu^+
\mu^-}}{c_S^{\,uc}}\mid^2 \,,
\end{eqnarray}
\begin{eqnarray}
\mathcal {B}^{\mathcal {U}}_{D^0\rightarrow e^+
e^-}\lesssim1.1\times10^{-23}\mid \frac{c_S^{\,e^+
e^-}}{c_S^{\,uc}}\mid^2 \,,
\end{eqnarray}
\begin{eqnarray}
\mathcal {B}^{\mathcal {U}}_{D^0\rightarrow \mu^+
e^-}\lesssim2.4\times10^{-19}\mid \frac{c_S^{\,\mu^+
e^-}}{c_S^{\,uc}}\mid^2 \,.
\end{eqnarray}
As is well recognized, due to the large experimental errors, only the order of magnitude of
these theoretical evaluations are meaningful.

The lagrangian determines
that $l(q)$ can be equal or unequal to $l'(q')$, thus, it is natural
to assume the couplings to be universal, namely a coupling takes a
value for all the same flavors and another value for all different
flavors, as discussed in Ref.~\cite{Ding:2008zza},
\begin{eqnarray}
c_S^{\,f' f}= \bigg \{ \begin{array}{cc}
  c_S^{}\,,\,\, & f \neq f' \\
  \kappa c_S^{}\,,\, & f = f'\, ,
\end{array}
\end{eqnarray}
where $\kappa>$1. To estimate the branching ratios, $\kappa=3$ is taken
as suggested by the authors of Ref.~\cite{Ding:2008zza}. The branching ratios $\mathcal
{B}^{\mathcal {U}}_{D^0\rightarrow l'^+ l^-}$ are
\begin{eqnarray}
\mathcal {B}^{\mathcal {U}}_{D^0\rightarrow \mu^+
\mu^-}\lesssim4.3\times10^{-18} \,,
\end{eqnarray}
\begin{eqnarray}
\mathcal {B}^{\mathcal {U}}_{D^0\rightarrow e^+
e^-}\lesssim1.0\times10^{-22} \,,
\end{eqnarray}
\begin{eqnarray}
\mathcal {B}^{\mathcal {U}}_{D^0\rightarrow \mu^+
e^-}\lesssim2.4\times10^{-19} \,.
\end{eqnarray}

\subsection{Non-universal $Z'$}

Next let us turn to the non-universal $Z'$ contribution to the decays
$D^0\rightarrow l'^+ l^-$. Taking $m_{Z'}\sim500$ GeV \cite{He:2007iu}, and just accounting the contributions from
non-universal $Z'$, with $\tan \theta_R\sim0.088$, the branching
ratios $\mathcal {B}_{D^0\rightarrow \mu^+ \mu^-,e^+ e^-}$ are
\begin{eqnarray}
\mathcal {B}_{D^0\rightarrow \mu^+ \mu^-}\lesssim 3.4\times 10^{-15}
\,,
\end{eqnarray}
\begin{eqnarray}
\mathcal {B}_{D^0\rightarrow e^+ e^-}\lesssim 7.9\times 10^{-20}\,.
\end{eqnarray}
For the lepton flavor violation case, we take the bound given in
Ref.~\cite{Chiang:2011cv} for our discussions. That is
\begin{eqnarray}
\mid \frac{b_R^{e \mu}}{m_{Z'}}\mid\lesssim1.8\times10^{-7}\,,
\end{eqnarray}
in unit of GeV$^{-1}$. The constraint is
\begin{eqnarray}
\mid \frac{g}{2 m_{Z'}} \tan \theta_W \cot \theta_R
V^{l^c\ast}_{R\,\tau e} V^{l^c}_{R\,\tau
\mu}\mid\lesssim1.8\times10^{-7}\,,
\end{eqnarray}
or
\begin{eqnarray}
\mid V^{l^c\ast}_{R\,\tau e} V^{l^c}_{R\,\tau
\mu}\mid\lesssim\frac{1}{\sqrt{\sqrt{2}G_F}}\times1.8\times10^{-7}\,.
\end{eqnarray}
The branching ratio $\mathcal {B}_{D^0\rightarrow \mu^+ e^-}$ is
\begin{eqnarray}
\mathcal {B}_{D^0\rightarrow \mu^+ e^-}\lesssim 5.5\times 10^{-20}
\,.
\end{eqnarray}

\section{The $D^0\rightarrow l'^+ l^-$ decay search  at BESIII and future charm-tau factory}

Since the first effort on limiting the branching fraction of FCNC
process $D^0\to\mu^+\mu^-$ was carried out by the European Muon
Collaboration \cite{Aubert:1985eg} in 1985, there have been many
experimental groups searchings for $D^0\to\mu^+\mu^-$,
$D^0\to\mu^{\pm}e^{\mp}$, and $D^0\to e^+e^-$ during the past thirty
years. Table \ref{tab::historical} summarizes their results, where
the $1^{\rm st}$ column refers the name of the experiments; the
$2^{\rm nd}$ column is for the year when the results were published;
the $3^{\rm rd}$ to $5^{\rm th}$ columns present the Upper Limit of
the branching fractions; the $6^{\rm th}$ column shows the
experiment style, {\it i.e.} fixed target, leptonic collider,
hadronic collider, or heavy ion collider; and the last two columns
correspond to the center-of-mass energies and data samples in use.
Most of the measurements suffered from high background
contaminations, and so the detection efficiency is rather low. The
important task for gaining meaningful conclusion is to enhance the
ability of distinguishing background and signal events. While, in
the experiments whose center-of-mass energy is near the $D^0\bar
D^0$ threshold, the neutral charm mesons are produced in pairs, one
can measure the di-lepton decays absolutely based on a technical
treatment namely double tagging method (i.e. to properly reconstruct
double $D$ mesons).
In the $e^+e^-$ annihilation experiment around 3.773 GeV, which is
just above the $D\bar D$ production threshold, $D\bar D$ pair is
produced via a decay of the resonance $\psi(3770)$ ($\psi(3770)\to
D\bar D$). If we only identify a fully reconstructed $\bar D$ meson
in one event, called as a singly tagged $\bar D$ meson, there must
exist a $D$ meson at the recoiling side. And if we reconstructed the
whole $D\bar D$ pair in the analysis procedure, the event will be
called as a doubly tagged event. Thus, with the data sample
consisting of the identified singly tagged $\bar D^0$ events, the
di-leptonic final states from decay of neutral $D$ mesons can be indubitably
selected, and the absolute branching fractions would be well
measured.
The advantage of the double tagging method  can extremely reduce
the background by tagging the $D$ meson pairs.
%
Historically, there were only two measurements of $D^0\to e^+e^-$
and $D^0\to\mu^{\pm}e^{\mp}$ using the threshold data by the MARK3
Collaboration, while they proceeded the analysis with single tagging
method (i.e. reconstruct only one $D$ meson) with a large
background, the threshold data did not bring up any advantages at
all.
Till now, the BESIII collaboration has accumulated 2.92
fb$^{-1}$ \cite{Ablikim:2012pj} $\psi(3770)$ data samples near its
production threshold during 11 month's data taking. There is about
$2.15\times10^7$ neutral $D$ mesons among $3.84\times10^7$ $D$
mesons assuming $\sigma^{\rm obs}_{D\bar D}=6.57$ nb
\cite{Dobbs:2007ab}. And we can eventually have more than 20
fb$^{-1}$ $\psi(3770)$ data according to the data taking plan of the
experiment, resulting a $D^0$ sample of about $1.47\times10^8$.
Then, the key issue will be, how many singly tagged $\bar D^0$
events we can reconstruct, and how well we can carry out the
measurement. To answer this question, here we present a full
simulation of searching for di-leptonic decays at the BESIII
experiment with the Monte Carlo method to discuss the experimental
sensitivities that can be reached in the future.

\begin{table*}[ht]
\tiny
\begin{center}
\caption{Historical measurements on searching for dilepton decays.} \label{tab::historical}
\begin{tabular}{cccccccc}
\hline Experiment & Year & $D^0\to \mu^+\mu^-$$[\times10^{-6}]$ & $D^0\to\mu^{\pm}e^{\mp}$$[\times10^{-6}]$ & $D^0\to e^+e^-$$[\times10^{-6}]$ & Style & Energy & Note \\
\hline
EMC\cite{Aubert:1985eg}      & 1985 & 340   & -     & -     & $\mu^- N$   & 280 GeV        & $1.3\times10^{12}$ Events \\
E615\cite{Biino:1985qc}    & 1986 & 11    & -     & -     & $\pi^- W$   & 225 GeV        & Norm. to $D$ decay  \\
MARK2\cite{Riles:1986jg}  & 1987 & -     & 2100  & -     & $e^+e^-$    & 29 GeV         &   \\
ACCMOR\cite{Palka:1987kx}  & 1987 & -     & 900   & -     & $\pi p$     & 200 GeV        &   \\
MARK3\cite{Becker:1987mu}  & 1987 & -     & 120   & -     & $e^+e^-$    & 3.77 GeV       & 9.3 pb$^{-1}$  \\
CLEO\cite{ref::CLEO-1988}    & 1988 & -     & 270   & 220   & $e^+e^-$    & 10 GeV         &   \\
ARGUS\cite{Albrecht:1988ge}  & 1988 & 70    & 100   & 170   & $e^+e^-$    & 10 GeV         &   \\
MARK3\cite{Adler:1987cp}   & 1988 & -     & -     & 130   & $e^+e^-$    & 3.77 GeV       & 9.6 pb$^{-1}$  \\
E789\cite{Mishra:1994ne}    & 1994 & 31    & -     & -     & $p N$       & -              &   \\
E653\cite{Kodama:1995ia}    & 1995 & 44    & -     & -     & $\pi^-$ emulsion & 600 GeV   &   \\
BEATRICE\cite{Adamovich:1995tp} & 1995 & 7.6  & -   & -    & $\pi^-C_u$  & 350 GeV        &   \\
CLEO2\cite{Freyberger:1996it}  & 1996 & 34    & 19    & 13    & $e^+e^-$    & $\Upsilon(4S)$ & 3.85 fb$^{-1}$  \\
E771\cite{Alexopoulos:1994hp}    & 1996 & 4.2   & -    & -    & $pS_i$      & 800 GeV        &   \\
BEATRICE\cite{Adamovich:1997wf} & 1997 & 4.1   & -    & -    & $\pi^-C_u$  & 350 GeV        &   \\
E791\cite{Aitala:1999db}    & 1999 & 5.2   & 8.1   & 6.2   & $\pi^- N$   & 500 GeV        & $2\times10^{10}$ Events \\
E789\cite{Pripstein:1999tq}    & 2000 & 15.6  & 17.2  & 8.19  & $p N$       & 800 GeV        & Norm. to $D^0\to K\pi$  \\
CDF\cite{Acosta:2003ag}      & 2003 & 2.5   & -     & -     & $p\bar p$   & 1.96 TeV       &   \\
BABAR\cite{Aubert:2004bs}  & 2004 & 2.0   & 0.81  & 1.2   & $e^+e^-$    & $\Upsilon(4S)$ & 122 fb$^{-1}$ \\
HERA-B\cite{Abt:2004hn} & 2004 & 2.0   & -     & -     & $pA$        & 920 GeV        &   \\
BELLE\cite{Petric:2010yt}  & 2010 & 0.14  & 0.26  & 0.079 & $e^+e^-$    & $\Upsilon(4S)$ & 660 fb$^{-1}$  \\
CDF\cite{Aaltonen:2010hz}      & 2010 & 0.21  & -     & -     & $p\bar p$   & 1.96 TeV       &   \\
LHCb\cite{Aaij:2013cza}    & 2013 & 0.0062  & -  & -  & $p p$    &  -     & 0.9 fb$^{-1}$  \\
BABAR\cite{Lees:2012jt}  & 2012 & 0.81  & 0.33  & 0.17  & $e^+e^-$    & 10.58 GeV      & 468 fb$^{-1}$  \\
PDG\cite{Beringer:1900zz}      & 2012 & 0.14  & 0.26  & 0.079 & -           & -              & - \\
\hline
\end{tabular}
\end{center}
\end{table*}

The Monte Carlo samples are obtained with the BESIII offline
Software System \cite{ref::boss}, where the particle trajectories
are simulated with a GEANT4 \cite{Agostinelli:2002hh} based package
\cite{ref::simbes} for the BESIII detector \cite{Ablikim:2009aa} at the
BEPC-II collider. The events used in this discussion, named as
generic MC events, are generated as $e^+e^-\to\psi(3770)\to
D\bar{D}$ at the c.m. energy $\sqrt{s}=$ 3.773 GeV with the $D\bar
D$ mesons decaying into all possible final states with the branching
fractions cited from PDG \cite{Beringer:1900zz}. Totally
$\sim1.31\times10^8$ $D\bar D$ events are produced at $\sqrt{s}=$
3.773 GeV, corresponding to an integrated luminosity of $\sim20$
fb$^{-1}$ $\psi(3770)$ data assuming $\sigma^{\rm obs}_{D\bar D} =
6.57$ nb \cite{Dobbs:2007ab}, which contains $\sim 7.35\times10^7$
$D^0\bar{D^0}$ pairs, $\sim6.08\times10^7$ $D^+D^-$ pairs.

The singly tagged $\bar D^0$ events are reconstructed in 4 golden
hadronic decays of $\bar D^0\to K^+\pi^-$(69\%), $\bar D^0\to
K^+\pi^-\pi^0$(35\%), $\bar D^0\to K^+\pi^-\pi^-\pi^+$(39\%), and
$\bar D^0\to K^+\pi^-\pi^-\pi^+\pi^0$(14\%), constituting more than
30\% of all $\bar D^0$ decays, where the numbers in brackets are
reconstruction efficiencies.
Tagged $\bar D^0$ events are filtered by two kinematic variables
based on the principles of energy and momentum conservations: (1)
Difference in energy $$\Delta E\equiv E_{\rm f}-E_{\rm b},$$ where
$E_{\rm f}$ is the total energy of the daughter particles from $\bar
D^0$ in one event and $E_{\rm b}$ is the $e^+/e^-$ beam energy for
the experiment, is recorded to describe the deviation from energy
conservation caused by experimental errors. (2) Beam-constrained
mass $$M_{\rm BC}\equiv\sqrt{E_{\rm b}^2-(\Sigma_i
\overrightarrow{p}_i)}$$ is calculated to reduce the uncertainty
caused by experimental errors when measuring the momenta of the
produced particles. In this definition, the energy $E_{\rm f}$ in
the expression of $$M^2_{\rm inv.}\equiv E^2_{\rm f}-p^2_{\rm f}$$
for the $\bar D$ invariant mass is replaced by $E_{\rm b}=E_{\rm
c.m.}/2$, where $E_{\rm c.m.}$ is the c.m. energy that $D^0\bar D^0$
produced.
The total energy and momentum of all the daughter particles in $\bar
D^0$ decays must satisfy the Energy Conservation (EC) principle,
generally one needs to introduce a kinematic fit, including energy
and momentum constraints and some correlated corrections, to reject
those not satisfying the EC which are caused by the uncertainty of
experimental measurement. This replacement of the real invariant
mass by $M_{\rm BC}$ partly plays the role.
Moreover, events are rejected if they fail to satisfy the selection
constraint $|\Delta E|<3\times\sigma_{\Delta E}$, which is tailored
for each individual decay mode, and $\sigma_{\Delta E}$ is the
standard deviation of the $\Delta E$ distribution.
If the $\bar D^0$ events were correctly tagged, a peak in $M_{\rm
BC}$ spectrum would emerge at the nominal mass of $\bar D^0$. Thus,
if there are more than one combinations in one tagged event, the one
with the smallest $|\Delta E|$ is retained. After considering the
detection efficiencies of each tag mode, $16856207\pm8874$ tagged
$\bar D^0$ events have been obtained based on simulated sample of about 20
fb$^{-1}$.

With the tagged $\bar D^0$ mesons, the $D^0$ decays into a lepton
pair is reconstructed in the recoiling side, i.e. $D^0\to
l^+l^{\prime-}$, where two charged track are identified as electrons
or muons. To suppress the contamination from $\gamma$ conversion,
the angle between electron and another charged tracks should be
greater than $30^{\circ}$.
And it is required that the $\Delta E$ distribution of the lepton
pairs should fall into the range of $|\Delta
E|<3\times\sigma_{\Delta E}$, where $\sigma_{\Delta E}$ is obtained
by fitting the $\Delta E$ distribution determined by the signal MC events. And
the valid signals would produce a peak at the $D^0$ nominal mass
within $3\sigma_{M_{\rm BC}}$ at the $M_{\rm BC}$ spectra.
For the processes of $D^0\to \mu^+\mu^-$, $D^0\to \mu^{\pm}e^{\mp}$,
and $D^0\to e^+e^-$, the numbers of estimated background events are
found to be all zero, by counting the signal window of
$|M_{e^+e^-}-M_{D^0}|<3\sigma_{M_{\rm BC}}$.

To examine the sensitivities of the measurement, we evaluate the
upper limits of the possible observed signal events, $s_{90}$, at
90\% confidence level, based on the expected background events
assuming zero signals. The upper limits are obtained by using the
Poissonian Limit Estimator (POLE) program \cite{Conrad:2002kn},
which is developed with an extended version \cite{Conrad:2002kn} of
the Feldman-Cousins method \cite{Feldman:1997qc}.
Thus, the upper limit on the branching fractions are calculated to
be
\begin{eqnarray}
&&\mathcal{B}(D^0\to \mu^+\mu^-)<4.7\times10^{-7}, \,\,
\mathcal{B}(D^0\to \mu^{\pm}e^{\mp})<3.4\times10^{-7},\nonumber\\
&&\mathcal{B}(D^0\to e^+e^-)<2.6\times10^{-7}
\end{eqnarray}
respectively, with
$$\mathcal{B}=\frac{s_{90}/\epsilon}{N^{\rm tag}_{\bar D^0}},$$ by
inserting the $s_{90}$, the detection efficiencies $\epsilon$(31\%
for $D^0\to\mu^+\mu^-$, 43\% for $D^0\to \mu^{\pm}e^{\mp}$, 55\% for
$D^0\to e^+e^-$ ), and the number of singly tagged $\bar D^0$ events
$N^{\rm tag}_{\bar D^0}$. The detection efficiencies are obtained by
analyzing the simulated events which are generated as $D^0\to
l^+l^{\prime-}$ and $\bar D^0\to anything$ with the same procedure
to the generic MC events.

The BEPCII collider is designed
to work at the c.m. energy of
$\sqrt{s}=3.773$ GeV with an instantaneous luminosity of $10^{33}$
${\rm cm^{-2}s^{-1}}$. As a conservative estimate, a data sample
with the integrated luminosity of about 20 fb$^{-1}$ can be
collected during less than 10 years' running.
This world largest threshold data sample will deliver an
experimental sensitivity for searching di-leptonic decays of $D^0$
meson of about $10^{-7}$ level. It seems that there will be a
desperate running time for the threshold experiment to challenge the
sensitivities from experiments at higher energies (e.g. $10^{-8}$ at
BELLE), however, it will not be a problem if one can have a
$\tau$-charm factory with an increasing of the luminosity of more
than 100 times.

\section{Conclusions}

In this article we give some discussions about the search of
flavor-changing interactions caused by new physics in $D^0$
leptonic decays. Considering the constraints set by the $D^0-\bar{D}^0$
mixing, we derive the new physics contributions: unparticle and
non-universal $Z'$ concerned in this work, to the decay modes $D^0\rightarrow\mu^+
\mu^-,e^+ e^-,\mu^{+} e^{-}$, and estimate the numerical results of
the rare decays $D^0\rightarrow l'^+ l^-$. The theoretical
predictions of branching ratios are shown in
Table~\ref{TabTh}, including contributions from SM and new physics from unparticle and
non-universal $Z'$.

\begin{table}
\begin{center}
\begin{tabular}{|c|c|c|c| }\hline\hline
            Branching ratios                                  & SM predictions & Unparticle &  Non-universal $Z'$  \\ \hline
            $\mathcal {B}_{D^0\rightarrow \mu^+ \mu^-}$       &  $10^{-13}$        & $\lesssim4.3\times10^{-18}$     & $\lesssim3.4\times10^{-15}$   \\
            $\mathcal {B}_{D^0\rightarrow e^+ e^-}$           & $10^{-23}$  & $\lesssim1.0\times10^{-22}$         &  $\lesssim7.9\times10^{-20}$  \\
            $\mathcal {B}_{D^0\rightarrow \mu^{\pm} e^{\mp}}$ & 0    & $\lesssim2.4\times10^{-19}$     &  $\lesssim5.5\times10^{-20}$   \\ \hline
\end{tabular}
\end{center}
\caption{The branching ratio predictions in $D^0\rightarrow l'^+
l^-$ decays, with the contributions from SM and new physics
unparticle, non-universal $Z'$.}\label{TabTh}
\end{table}

For the decay $D^0\rightarrow \mu^+ \mu^-$, it is shown that the long-distance
effect of SM still exceeds the contributions from  unparticle and
non-universal $Z'$, therefore the two models do not manifest in the decays.
But if the leptonic decay $D^0\rightarrow \mu^+ \mu^-$ is observed with larger branching ratio (larger that $10^{-13}$),
it indicates that there exist BSM contributions, but not from unparticle or non-universal $Z'$.
Since $D^0\rightarrow e^+ e^-$ suffers from the helicity suppression in the SM, so that the new physics contribution
may exceed the SM contribution, but this branching ratio is
very small to be observed with the present facilities. A simple analysis indicates that the decay mode $D^0\rightarrow
\mu^{\pm} e^{\mp}$ is much suppressed in SM. Therefore a sizable or at least observable mode
$D^0\rightarrow \mu^{\pm} e^{\mp}$ must be due to new physics
contributions.

As discussed in this work, even though the leptonic decays of $D^0$ are sensitive
to the new physics as implied by the measured $D^0-\bar D^0$ mixing, the contributions from unparticle and non-universal $Z'$ cannot exceed the SM contribution. The
favorable modes which may distinguish between the SM and BSM contributions are the lepton-flavor violation processes
which are much suppressed in the SM. However, the branching ratio of such modes are very small, even though some BSM mechanisms
such as unparticle and non-universal $Z'$ are taken into account. They are far below the reach of any presently available facilities. In fact there are many new physics models which might cause a larger branching ratio (other schemes, see e.g. \cite{Golowich:2009ii}). The measurement on
the leptonic decays $D^0\to\mu^+\mu^-$ is worthwhile and one might find a trace of new physics. Meanwhile
$D^0\to\mu^+e^-(\mu^-e^+)$ is a much better place to look for
new physics.

Even though the present facilities cannot provide large amount of $D^0$,
one may expect that the future super charm-tau factory and LHC may do the job.

\acknowledgments \vspace*{-3ex} This work was supported in part by
National Natural Science Foundation of China under the contract No. 11375128.


\begin{thebibliography}{0} \vspace*{-2ex}


\bibitem{Aubert:2007wf}
  B.~Aubert {\it et al.}  [BaBar Collaboration],
  Phys.\ Rev.\ Lett.\  {\bf 98} (2007) 211802  [hep-ex/0703020 [HEP-EX]].  

\bibitem{Staric:2007dt}
  M.~Staric {\it et al.}  [Belle Collaboration],
  Phys.\ Rev.\ Lett.\  {\bf 98} (2007) 211803  [hep-ex/0703036].  

\bibitem{Aaltonen:2007ac}
  T.~Aaltonen {\it et al.}  [CDF Collaboration],
  Phys.\ Rev.\ Lett.\  {\bf 100} (2008) 121802
  [arXiv:0712.1567 [hep-ex]].


\bibitem{Chen:2007yn}
  C.~-H.~Chen, C.~-Q.~Geng and T.~-C.~Yuan,
  Phys.\ Lett.\ B {\bf 655} (2007) 50
  [arXiv:0704.0601 [hep-ph]].


\bibitem{Hou:2006mx}
  W.~-S.~Hou, M.~Nagashima and A.~Soddu,
  Phys.\ Rev.\ D {\bf 76} (2007) 016004
  [hep-ph/0610385].


\bibitem{He:2007iu}
  X.~-G.~He and G.~Valencia,
  Phys.\ Lett.\ B {\bf 651} (2007) 135
  [hep-ph/0703270].


\bibitem{Li:2007by}
  X.~-Q.~Li and Z.~-T.~Wei,
  Phys.\ Lett.\ B {\bf 651} (2007) 380  [arXiv:0705.1821 [hep-ph]].



\bibitem{Aaij:2012nna}
  RAaij {\it et al.}  [LHCb Collaboration],
  Phys.\ Rev.\ Lett.\  {\bf 110} (2013) 021801
  [arXiv:1211.2674 [hep-ex]].

\bibitem{Aaij:2013aka}
  RAaij {\it et al.}  [LHCb Collaboration],
  Phys.\ Rev.\ Lett.\  {\bf 111} (2013) 101805
  [arXiv:1307.5024 [hep-ex]].

\bibitem{Chatrchyan:2013bka}
  S.~Chatrchyan {\it et al.}  [CMS Collaboration],
  Phys.\ Rev.\ Lett.\  {\bf 111} (2013) 101804
  [arXiv:1307.5025 [hep-ex]].


\bibitem{Gorn:1978sb}
  M.~Gorn,
  Phys.\ Rev.\ D {\bf 20}, 2380 (1979).

\bibitem{Pakvasa:1994ni}
  S.~Pakvasa,
  Chin.\ J.\ Phys.\  {\bf 32} (1994) 1163
  [hep-ph/9408270].

\bibitem{Burdman:2001tf}
  G.~Burdman, E.~Golowich, J.~L.~Hewett and S.~Pakvasa,
  Phys.\ Rev.\ D {\bf 66}, 014009 (2002)
  [hep-ph/0112235].


\bibitem{Georgi:2007ek}
  H.~Georgi,
  Phys.\ Rev.\ Lett.\  {\bf 98} (2007) 221601
  [hep-ph/0703260].


\bibitem{Barger:2003hg}
  V.~Barger, C.~-W.~Chiang, P.~Langacker and H.~-S.~Lee,
  Phys.\ Lett.\ B {\bf 580} (2004) 186
  [hep-ph/0310073].

\bibitem{He:2004it}
  X.~-G.~He and G.~Valencia,
  Phys.\ Rev.\ D {\bf 70}, 053003 (2004)
  [hep-ph/0404229].

\bibitem{Cheung:2006tm}
  K.~Cheung, C.~-W.~Chiang, N.~G.~Deshpande and J.~Jiang,
  Phys.\ Lett.\ B {\bf 652} (2007) 285
  [hep-ph/0604223].

\bibitem{He:2006bk}
  X.~-G.~He and G.~Valencia,
  Phys.\ Rev.\ D {\bf 74} (2006) 013011
  [hep-ph/0605202].

\bibitem{Chiang:2006we}
  C.~-W.~Chiang, N.~G.~Deshpande and J.~Jiang,
  JHEP {\bf 0608} (2006) 075
  [hep-ph/0606122].

\bibitem{Baek:2006bv}
  S.~Baek, J.~H.~Jeon and C.~S.~Kim,
  Phys.\ Lett.\ B {\bf 641} (2006) 183
  [hep-ph/0607113].


\bibitem{Li:2010af}
  H.~-B.~Li and M.~-Z.~Yang,
  Sci.\ China G {\bf 53} (2010) 1953.

\bibitem{Peng:2013wsa&Li:2013mnl}
  T.~Peng [Belle Collaboration],
  PoS ICHEP {\bf 2012} (2013) 357;
  L.~Li,
  arXiv:1310.6142 [hep-ex].

\bibitem{Aaij:2013wda}
  RAaij {\it et al.}  [LHCb Collaboration],
  arXiv:1309.6534 [hep-ex].


\bibitem{Aaij:2013ria}
  RAaij {\it et al.}  [ LHCb Collaboration],
  arXiv:1310.7201 [hep-ex].  



\bibitem{Georgi:2007si}
  H.~Georgi,
  Phys.\ Lett.\ B {\bf 650} (2007) 275
  [arXiv:0704.2457 [hep-ph]].


\bibitem{Cheung:2007zza}
  K.~Cheung, W.~-Y.~Keung and T.~-C.~Yuan,
  Phys.\ Rev.\ Lett.\  {\bf 99} (2007) 051803
  [arXiv:0704.2588 [hep-ph]].


\bibitem{Luo:2007bq}
  M.~Luo and G.~Zhu,
  Phys.\ Lett.\ B {\bf 659} (2008) 341
  [arXiv:0704.3532 [hep-ph]].

\bibitem{Grinstein:2008qk}
  B.~Grinstein, K.~A.~Intriligator and I.~Z.~Rothstein,
  Phys.\ Lett.\ B {\bf 662} (2008) 367
  [arXiv:0801.1140 [hep-ph]].

\bibitem{Chen:2007cz}
  S.~-L.~Chen, X.~-G.~He, X.~-Q.~Li, H.~-C.~Tsai and Z.~-T.~Wei,
  Eur.\ Phys.\ J.\ C {\bf 59} (2009) 899  [arXiv:0710.3663 [hep-ph]].

\bibitem{Burdman:2003rs}
  G.~Burdman and I.~Shipsey,
  Ann.\ Rev.\ Nucl.\ Part.\ Sci.\  {\bf 53} (2003) 431  [hep-ph/0310076].


\bibitem{He:2002ha}
  X.~-G.~He and G.~Valencia,
  Phys.\ Rev.\ D {\bf 66} (2002) 013004
   [Erratum-ibid.\ D {\bf 66} (2002) 079901]
  [hep-ph/0203036].

\bibitem{He:2003qv}
  X.~-G.~He and G.~Valencia,
  Phys.\ Rev.\ D {\bf 68} (2003) 033011
  [hep-ph/0304215].


\bibitem{Lee:2013dma}
  J.~-P.~Lee,
  Phys.\ Rev.\ D {\bf 88} (2013) 116003
  [arXiv:1303.4858 [hep-ph]].



\bibitem{Beringer:1900zz}
  J.~Beringer {\it et al.}  [Particle Data Group Collaboration],
  Phys.\ Rev.\ D {\bf 86} (2012) 010001.

\bibitem{Ding:2008zza}
  G.~-J.~Ding and M.~-L.~Yan,
  Phys.\ Rev.\ D {\bf 77} (2008) 014005.

\bibitem{Chiang:2011cv}
  C.~-W.~Chiang, Y.~-F.~Lin and J.~Tandean,
  JHEP {\bf 1111} (2011) 083
  [arXiv:1108.3969 [hep-ph]].


\bibitem{Aubert:1985eg}
  J.~J.~Aubert {\it et al.}  [European Muon Collaboration],
  Phys.\ Lett.\ B {\bf 155} (1985) 461.

\bibitem{Biino:1985qc}
  C.~Biino, J.~F.~Greenhalgh, W.~C.~Louis, K.~T.~McDonald, S.~Palestini, F.~C.~Shoemaker, A.~J.~S.~Smith and C.~E.~Adolphsen {\it et al.},
  Phys.\ Rev.\ Lett.\  {\bf 56} (1986) 1027.

\bibitem{Riles:1986jg}
  K.~Riles, J.~Dorfan, G.~S.~Abrams, D.~Amidei, A.~R.~Baden, T.~Barklow, A.~Boyarski and J.~Boyer {\it et al.},
  Phys.\ Rev.\ D {\bf 35} (1987) 2914.

\bibitem{Palka:1987kx}
  H.~Palka {\it et al.}  [ACCMOR Collaboration],
  Phys.\ Lett.\ B {\bf 189} (1987) 238.

\bibitem{Becker:1987mu}
  J.~Becker {\it et al.}  [MARK-III Collaboration],
  Phys.\ Lett.\ B {\bf 193} (1987) 147
   [Erratum-ibid.\ B {\bf 198} (1987) 590].


\bibitem{ref::CLEO-1988}
  P.~Haas {\it et al.}  [Cleo Collaboration],
  Phys.\ Rev.\ Lett.\  {\bf 60} (1988) 1614.

\bibitem{Albrecht:1988ge}
  H.~Albrecht {\it et al.}  [ARGUS Collaboration],
  Phys.\ Lett.\ B {\bf 209} (1988) 380.

\bibitem{Adler:1987cp}
  J.~Adler {\it et al.}  [MARK-III Collaboration],
  Phys.\ Rev.\ D {\bf 37} (1988) 2023
   [Erratum-ibid.\ D {\bf 40} (1989) 3788].

\bibitem{Mishra:1994ne}
  C.~S.~Mishra {\it et al.}  [E789 Collaboration],
  Phys.\ Rev.\ D {\bf 50} (1994) 9.

\bibitem{Kodama:1995ia}
  K.~Kodama {\it et al.}  [E653 Collaboration],
  Phys.\ Lett.\ B {\bf 345} (1995) 85.

\bibitem{Adamovich:1995tp}
  M.~Adamovich {\it et al.}  [BEATRICE Collaboration],
  Phys.\ Lett.\ B {\bf 353} (1995) 563.

\bibitem{Freyberger:1996it}
  A.~Freyberger {\it et al.}  [CLEO Collaboration],
  Phys.\ Rev.\ Lett.\  {\bf 76} (1996) 3065
   [Erratum-ibid.\  {\bf 77} (1996) 2147].

\bibitem{Alexopoulos:1994hp}
  T.~Alexopoulos {\it et al.}  [E771 Collaboration],
  Phys.\ Rev.\ Lett.\  {\bf 77} (1996) 2380.

\bibitem{Adamovich:1997wf}
  M.~Adamovich {\it et al.}  [BEATRICE Collaboration],
  Phys.\ Lett.\ B {\bf 408} (1997) 469.

\bibitem{Aitala:1999db}
  E.~M.~Aitala {\it et al.}  [E791 Collaboration],
  Phys.\ Lett.\ B {\bf 462} (1999) 401
  [hep-ex/9906045].

\bibitem{Pripstein:1999tq}
  D.~Pripstein {\it et al.}  [E789 Collaboration],
  Phys.\ Rev.\ D {\bf 61} (2000) 032005
  [hep-ex/9906022].

\bibitem{Acosta:2003ag}
  D.~Acosta {\it et al.}  [CDF Collaboration],
  Phys.\ Rev.\ D {\bf 68} (2003) 091101
  [hep-ex/0308059].

\bibitem{Aubert:2004bs}
  B.~Aubert {\it et al.}  [BaBar Collaboration],
  Phys.\ Rev.\ Lett.\  {\bf 93} (2004) 191801
  [hep-ex/0408023].

\bibitem{Abt:2004hn}
  I.~Abt {\it et al.}  [HERA-B Collaboration],
  Phys.\ Lett.\ B {\bf 596} (2004) 173
  [hep-ex/0405059].

\bibitem{Petric:2010yt}
  M.~Petric {\it et al.}  [Belle Collaboration],
  Phys.\ Rev.\ D {\bf 81} (2010) 091102
  [arXiv:1003.2345 [hep-ex]].

\bibitem{Aaltonen:2010hz}
  T.~Aaltonen {\it et al.}  [CDF Collaboration],
  Phys.\ Rev.\ D {\bf 82} (2010) 091105
  [arXiv:1008.5077 [hep-ex]].

\bibitem{Aaij:2013cza}
  R.~Aaij {\it et al.}  [LHCb Collaboration],
  Phys.\ Lett.\ B {\bf 725} (2013) 15
  [arXiv:1305.5059 [hep-ex]].
  

\bibitem{Lees:2012jt}
  J.~P.~Lees {\it et al.}  [BaBar Collaboration],
  Phys.\ Rev.\ D {\bf 86} (2012) 032001
  [arXiv:1206.5419 [hep-ex]].



\bibitem{Ablikim:2012pj}
  M.~Ablikim {\it et al.}  [BESIII Collaboration],
  Chin.\ Phys.\ C {\bf 37} (2013) 063001
  [arXiv:1209.6199 [hep-ex]].

\bibitem{Dobbs:2007ab}
  S.~Dobbs {\it et al.}  [CLEO Collaboration],
  Phys.\ Rev.\ D {\bf 76} (2007) 112001
  [arXiv:0709.3783 [hep-ex]].

\bibitem{ref::boss} W. D. Li {\it et al.}, The Offine Software for the BESIII Experiment, Proceeding of CHEP06 (Mumbai, India, February 2006)

\bibitem{Agostinelli:2002hh}
  S.~Agostinelli {\it et al.}  [GEANT4 Collaboration],
  Nucl.\ Instrum.\ Meth.\ A {\bf 506} (2003) 250.


\bibitem{ref::simbes}
Zi-Yan Deng, Guo-Fu Cao, Cheng-Dong Fu, etc., High Energy Physics and Nuclear Physics, 2006, {\bf 30} (05): 371-377 (in Chinese).


\bibitem{Ablikim:2009aa}
  M.~Ablikim {\it et al.}  [BESIII Collaboration],
  Nucl.\ Instrum.\ Meth.\ A {\bf 614} (2010) 345
  [arXiv:0911.4960 [physics.ins-det]].




\bibitem{Conrad:2002kn}
  J.~Conrad, O.~Botner, A.~Hallgren and C.~Perez de los Heros,
  Phys.\ Rev.\ D {\bf 67} (2003) 012002
  [hep-ex/0202013].


\bibitem{Feldman:1997qc}
  G.~J.~Feldman and R.~D.~Cousins,
  Phys.\ Rev.\ D {\bf 57} (1998) 3873
  [physics/9711021 [physics.data-an]].

\bibitem{Golowich:2009ii}
  E.~Golowich, J.~Hewett, S.~Pakvasa and A.~A.~Petrov,
  Phys.\ Rev.\ D {\bf 79} (2009) 114030
  [arXiv:0903.2830 [hep-ph]].





\end{thebibliography}
\end{document}